\documentclass[journal=jacsat,manuscript=article]{achemso}

\usepackage{chemformula} 
\usepackage[T1]{fontenc} 
\usepackage{graphicx}
\usepackage{float}
\usepackage{epsfig}
\usepackage{gensymb}
\usepackage{latexsym}
\usepackage{amsmath}
\usepackage{xfrac}
\usepackage{amssymb}
\usepackage{xspace}
\usepackage{caption}
\usepackage{subcaption}
\usepackage{wrapfig}
\usepackage{xpatch}
\usepackage{xcolor}
\usepackage{array}
\usepackage[export]{adjustbox}

\author{B. Gorin}
\affiliation[University of Amsterdam]
{Van der Waals Zeeman Institute, University of Amsterdam,1018 XE Amsterdam, The Netherlands}
\alsoaffiliation[University of Bordeaux]
{Laboratoire Ondes et Matière d'Aquitaine, Université de Bordeaux, 33400 Talence, France}
\email{benjamin.gorin@u-bordeaux.fr}
\author{G. Di Mauro}
\affiliation[University of Amsterdam]
{Van der Waals Zeeman Institute, University of Amsterdam,1018 XE Amsterdam, The Netherlands}
\email{gabrielle.di-mauro@espci.fr}
\author{D. Bonn}
\affiliation[University of Amsterdam]
{Van der Waals Zeeman Institute, University of Amsterdam,1018 XE Amsterdam, The Netherlands}
\email{D.Bonn@uva.nl}
\author{H. Kellay}
\affiliation[University of Bordeaux]
{Laboratoire Ondes et Matière d'Aquitaine, Université de Bordeaux, 33400 Talence, France}
\alsoaffiliation[University of Bordeaux]
{Institut Universitaire de France, 75005 Paris, France}
\email{hamid.kellay@u-bordeaux.fr}

\title[An \textsf{achemso} demo]
  {Universal aspects of droplet spreading dynamics in Newtonian and non-Newtonian Fluids}

\begin{document}







\begin{abstract}
Droplet impacts are common in many applications such as coating, spraying, or printing; understanding how droplets spread after impact is thus of utmost importance. Such impacts may occur with different velocities on a variety of substrates. The fluids may also be non-Newtonian and thus possess different rheological properties. How the different properties such as surface roughness and wettability, droplet viscosity and rheology as well as interfacial properties affect the spreading dynamics of the droplets and the eventual drop size after impact are unresolved questions. 

Most recent work focuses on the maximum spreading diameter after impact and uses scaling laws to predict this. In this paper we show that a proper rescaling of the spreading dynamics with the maximum radius attained by the drop, and the impact velocity leads to a unique single and thus universal curve for the variation of diameter versus time. The validity of this universal functional shape is validated for different liquids with different rheological properties as well as substrates with different wettabilities. This universal function agrees with a recent model that proposes a closed set of differential equations for the spreading dynamics of droplets. 
\end{abstract}

\section{Introduction}

Droplet impacts are routinely studied \cite{josserand_drop_2016} to understand a wide range of phenomena ranging from spraying, coating, printing \cite{hoath2016fundamentals,glasser_tuning_2019,xie_delamination_2021,lohse_fundamental_2022} to self cleaning and anti-icing \cite{thievenaz_solidification_2019}. Typical examples abound from every day life or in industrial applications: a droplet of rain impacting a glass pane \cite{cebeci_aircraft_2003}, the deposition of a molten metal droplet for electronic printing \cite{fukanuma_porosity_1994,tavakoli_spreading_2014,gielen_solidification_2020}, or droplet impacts of polymers or surfactants solutions usually used for a more controlled deposition \cite{bergeron_controlling_2000,bartolo_retraction_2005,bartolo_dynamics_2007,jalaal_gel-controlled_2018,hoffman_controlling_2021}. It is thus central to understand droplet spreading and give it a proper description. 
The difficulty of understanding droplet impacts stems from the fact that this process involves many different contributions such as viscous dissipation, surface tension effects, surface roughness, wetting and spreading as well as the dynamics of the contact line. It is therefore arduous to propose a simple description for the spreading dynamics upon impact and describe how the diameter of the drop $D(t)$ evolves with time $t$ \cite{biance_first_2004,eggers_drop_2010,lagubeau_spreading_2012,eddi_short_2013,wildeman_spreading_2016} .
Many studies focus on the maximum spreading diameter $D_{max}$ and propose different scaling laws using arguments of energy conservation \cite{madejski_droplets_1983,clanet_maximal_2004,ukiwe_maximum_2005,laan_maximum_2014,lee_modeling_2016,wei_maximum_2021} .
Recently, for the purpose of describing the dynamics of spreading of Newtonian fluids, a theory proposed by Gordillo et.al  \cite{riboux_maximum_2016,gordillo_theory_2019,garcia-geijo_spreading_2021} gives a solution for $D(t)$ by solving a system of equations describing the spreading of an impacting droplet. This model is based on modelling the spreading droplet as a thick rim followed by a thin liquid film connected to the bulk of the drop. Using mass and momentum conservation between the rim, the liquid film and the droplet, a set of coupled differential equations is proposed.
Solving these equations with the proper boundary conditions leads to a complete description of $D(t)$ for Newtonian drops impacting a smooth surface. In this work, we carry out experiments of droplet spreading dynamics on a variety of fluids, with different rheological properties, impacting at different velocities substrates with different wettabilities. Our results show that a simple rescaling of $D(t)$ with the maximum diameter $D_{max}$ and a rescaling of the time scale by a characteristic time $D_{max}/V_0$ where $V_0$ is the impact velocity, leads to a single master curve describing the spreading dynamics of  droplet impacts independently of the substrate, interfacial or rheological properties. Our scaling function agrees with the recent theory of Gordillo et.al \cite{gordillo_theory_2019} . Our work therefore shows that the diameter $D(t)$ of spreading droplets depends only on the maximum spreading diameter $D_{max}$ and the impact velocity $V_0$. 

\section{Methods and materials}

\begin{figure}
     \centering
    \includegraphics[width=\textwidth]{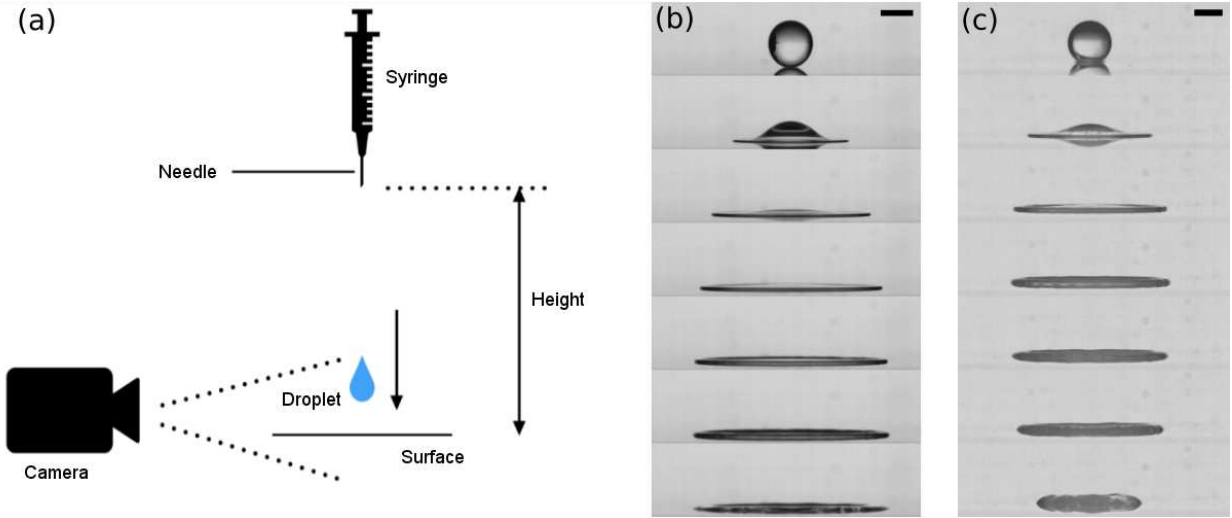}
    \caption{(a) experimental setup for measuring droplet impacts on a flat surface. (b) water droplet ($D_{0}=3.57 \; mm$, $V_{0,a}=2.29 \; m.s^{-1}$, $Re=8100$, $We=260$)  impacting a hydrophilic glass surface. (c) water droplet ($D_{0}=3.64 \; mm$, $V_{0}=2.17\; m.s^{-1}$, $Re=7900$, $We=240$) impacting a hydrophobic glass surface. Snapshots are taken at 0-1-2-3-4-5-8 ms after impact. Scale bar : 2 mm}
    \label{fig1}
\end{figure}

We perform a series of droplet impact experiments with different liquids 
(water, hexadecane, aqueous polymer solutions and Carbopol solutions) impacting different surfaces.  The experimental setup is as follows: a syringe with a needle is fixed at a certain height which sets the impact velocity $V_0$ 
(varied between $1m.s^{-1}$ and $4m.s^{-1}$).
The syringe is connected to a syringe pump to release droplets using very low injection fluxes. Droplet sizes $D_0$ can be varied between $1.80$ and $4.5 mm$ (all droplet diameters are given in the figure descriptions). 
The Reynolds and Weber numbers are defined as $Re=\frac{\rho V_0 D_0}{\eta}$ and $We=\frac{\rho V_0^2 D_0}{\sigma}$ where $\rho$, $\eta$ and $\sigma$ are respectively the density, the viscosity and the surface tension of the liquid. Different surfaces are used: smooth hydrophilic and hydrophobic microscope glass slides (hydrophobic coating with Octadecyltrichlorosilane) as well as parafilm. 
All experiments are done at room temperature $T=23 \degree C$ and $RH=40 \%$. 
Droplet impacts are recorded at high framerates (up to 19000 fps) using a fast camera (Phantom V640) aligned with the substrate ( see figure \ref{fig1}a,b and c).

\makeatletter
\let\@float@original\@float
\xpatchcmd{\@float}{\csname fps@#1\endcsname}{ht!}{}{}
\makeatother
\makeatletter
\let\@float\@float@original
\makeatother

The liquids used for the droplets are pure water, hexadecane and two types of non-Newtonian fluids. We use aqueous solutions of polymer (polyethylene oxide also called PEO) of molecular weight of $4.10^6g.mol^{-1}$ at concentrations varying between $10 \; ppm$ and $5000 \; ppm$. A small amount of isopropanol (about 1 wt \%) is added to avoid polymer aggregation. Experiments are done using a yield stress fluid consisting of aqueous solutions of Carbopol at different concentrations. 
In all cases, we focus our study on drop impacts with high Reynolds numbers meaning that the spreading is mainly driven by droplet inertia. Further, drop impact velocities are chosen in such a way that splashing is avoided.
The two Non-Newtonian fluids are used in order to test the role of the liquid rheology.
The presence of polymer renders the viscosity of the liquid shear rate dependent and elastic polymer flow interactions can take place \cite{ingremeau_stretching_2013,bonn_viscoelastic_1997}. Further experiments using Carbopol solutions are intended to examine how yield stress fluids affect the spreading \cite{louvet_nonuniversality_2014,martouzet_dynamic_2021} . 

The rheology of the polymer solutions (PEO) was measured using an Anton Paar rheometer with a cone plate geometry (two cones were used, angle $0.5\degree$ and diameter 60 $mm$ and angle $1\degree$ and diameter 50$mm$). For the Carbopol solutions the flow curves were measured using a rough cone-plate geometry (diameter = 40 $mm$ , angle = $1\degree58'$ , truncation
gap= 59 $\mu m$) on a controlled stress rheometer ARG2 (TA - Instruments). 
The relaxation time for the 5000 ppm PEO aqueous solutions using the Carreau Model is $\tau_r \sim 0.040 \; s$. For Carbopol solutions, we use the same definition as proposed by Luu.et.al \cite{luu_drop_2009} to estimate a characteristic time as : $\tau_r=(\frac{k}{G})^{1/n}$ where $k$ is the fluid consistency (in $Pa.s^{n} $), G the elastic shear modulus (in $Pa.$) and $n$ the flow index.
\begin{figure}
     \centering
    \includegraphics[width=0.95\textwidth]{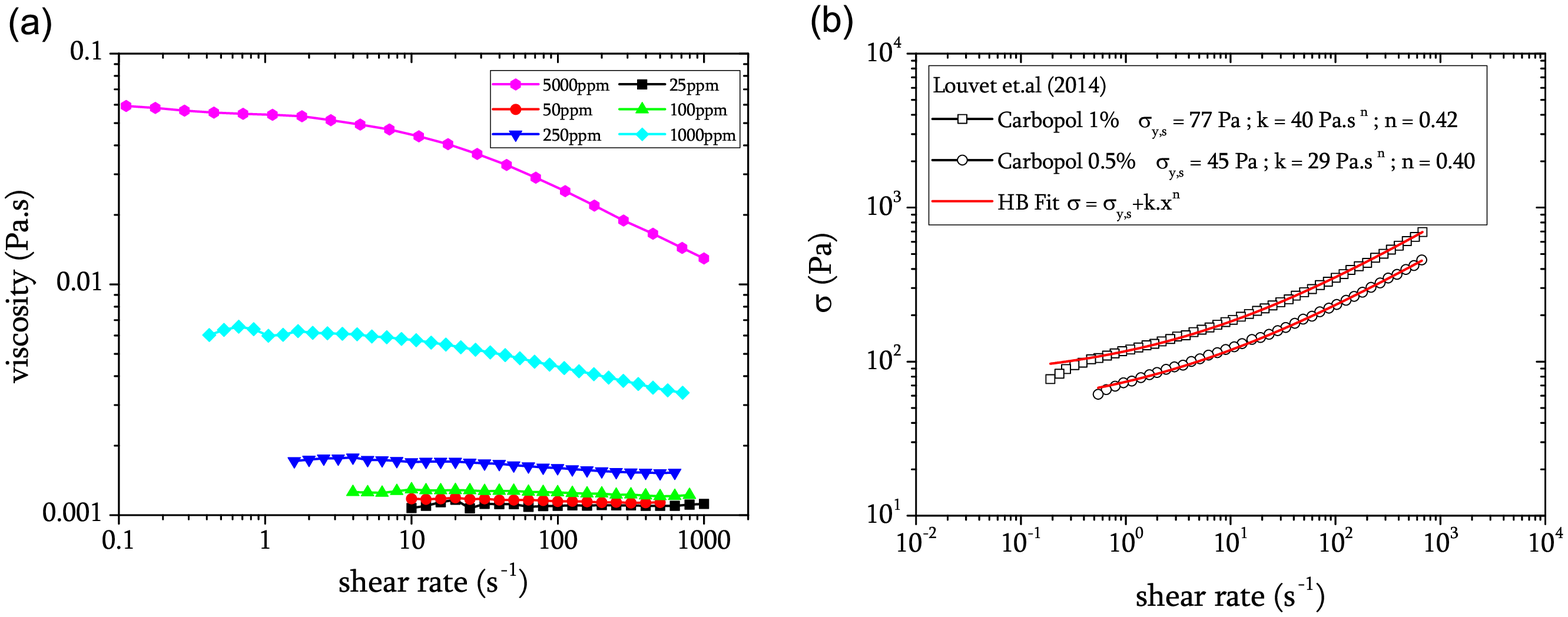}
    \caption{Rheology of (a) PEO at different concentrations and (b) Carbopol solutions at two different concentrations. The flow curves are fit to the Hershel Bulkley model $\sigma=\sigma_{ys}+kx^n$. Here $x$ is the shear rate, $n$ the flow index, $k$ the consistency parameter, and $\sigma_{ys}$ is the yield stress of the solution. The values of the parameters are given in the figure.}
    \label{fig2}
\end{figure}
The Deborah number is defined as the ratio of the relaxation time $\tau_r$ deduced from the rheology of the system and the observation time $\tau=\frac{R_0}{V_0}$: $De=\frac{\tau_r}{\tau}$. For all impacts $De$ was much greater than 1 (10 for the Carbopol solutions and 40 for the 5000ppm PEO solution at a velocity of 1m/s) so our drops behave like a fluid during the impact.

In general, when a droplet impacts a flat smooth surface, it follows three different regimes : An initial stage where the droplet spreads very quickly until it reaches a maximum spreading diameter $D_{max}$. In principle, during this stage, surface wettability is negligible compared to droplet inertia. When the droplet reaches its maximum spreading diameter, it enters a stage where wetting plays a role and the droplet can either retract or continue spreading \cite{tanner_spreading_1979} depending on the surface wettability. 

\makeatletter
\let\@float@original\@float
\xpatchcmd{\@float}{\csname fps@#1\endcsname}{ht!}{}{}
\makeatother
\begin{figure}
     \centering
    \includegraphics[width=\textwidth]{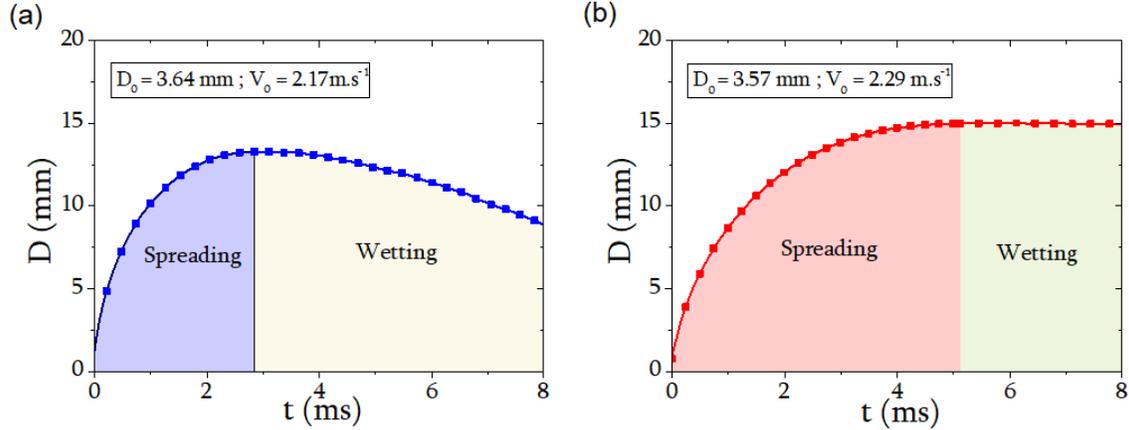}
        \caption{Spreading diameter dynamics $D(t)$ of a water drop of figure \ref{fig1} impacting a (a) hydrophobic and (b) hydrophilic microscope glass slide}
        \label{fig3}
\end{figure}
\makeatletter
\let\@float\@float@original
\makeatother

In the work carried out here, most droplets retract because the impact velocity is sufficiently high that the maximum spreading diameter is always larger than the equilibrium spreading diameter $D_{max} > D_{eq}$. The retraction speed then depends on the wettability of the surface and the liquid properties.

This phenomenology is illustrated in figure \ref{fig3} where two experimental examples of the spreading dynamics $D(t)$ of droplets impacting different substrates are shown. In the first case little retraction is observed as the surface is hydrophilic. When the surface is highly hydrophobic, retraction occurs as in the second example. This retraction can lead to a possible rebound of the droplet. Here we only focus on the spreading dynamics until the droplets reach the maximum spreading diameter $D_{max}$. 

\section{Results and discussion}

\begin{figure}
    \centering
    \includegraphics[width=\textwidth]{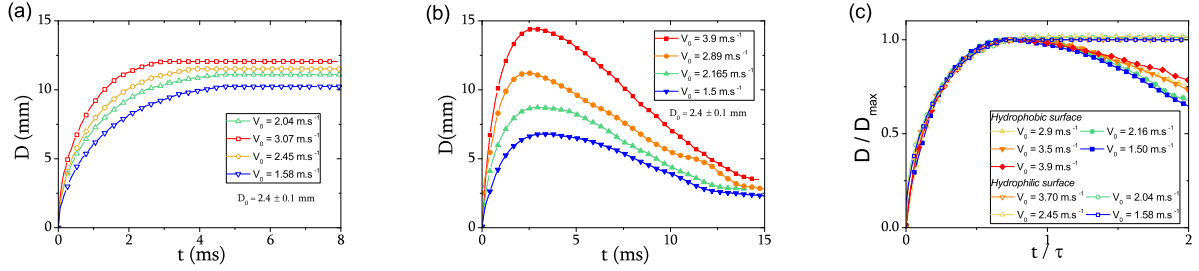}
    \caption{Spreading diameter dynamics of water droplets impacting two different substrates : (a) a hydrophilic glass and (b) a hydrophobic glass. (c) Rescaled spreading dynamic of water drop impacts. The Reynolds and Weber numbers varied between $3500<Re<8700$ and $77<We<470$}
    \label{fig4}
\end{figure}

We first discuss the results for water drops impacting a hydrophilic and a hydrophobic glass substrate. This is shown in figures \ref{fig4}a and b.
For both cases, the spreading is fast in the initial stages, reaches a maximum diameter, before settling on a quasi plateau value, figure \ref{fig4}a, or before retracting for the hydrophobic substrate, figure \ref{fig4}b. Note that the maximum spreading diameter as well as the speed of spreading increase as the velocity of impact increases as expected for both substrates. Our main result is shown in figure \ref{fig4}c. Here we replot the same data as figures \ref{fig4}a and \ref{fig4}b using rescaled variables. The time of evolution is rescaled as $t/\tau$ with $\tau=D_{max}/V_0$ while the diameter is rescaled as $D/D_{max}$. This figure shows that the spreading diameter dynamics follows one single universal functional shape until it reaches its maximum value $D_{max}$ with deviations occurring at later times. At these late times, either retraction occurs (in particular for the hydrophobic surface) or the droplet continues to spread slowly due to surface wettability. In these regimes, the collapse of the data ceases to apply.

\makeatletter
\let\@float@original\@float
\xpatchcmd{\@float}{\csname fps@#1\endcsname}{ht!}{}{}
\makeatother
\makeatletter
\let\@float\@float@original
\makeatother

In fact this scaling function, for the spreading dynamics up to the moment where the maximum diameter is reached, is insensitive to the wettability of the substrate as the collapse of the data from the two substrates show. This is our main result and show that a proper rescaling of the diameter and the time scale collapses our data for different substrates, different initial diameters as well as different impact velocities. Does this rescaling persist for other fluids and rheological properties? This is what we examine next. 

\begin{figure}
     \centering
    \includegraphics[width=\textwidth]{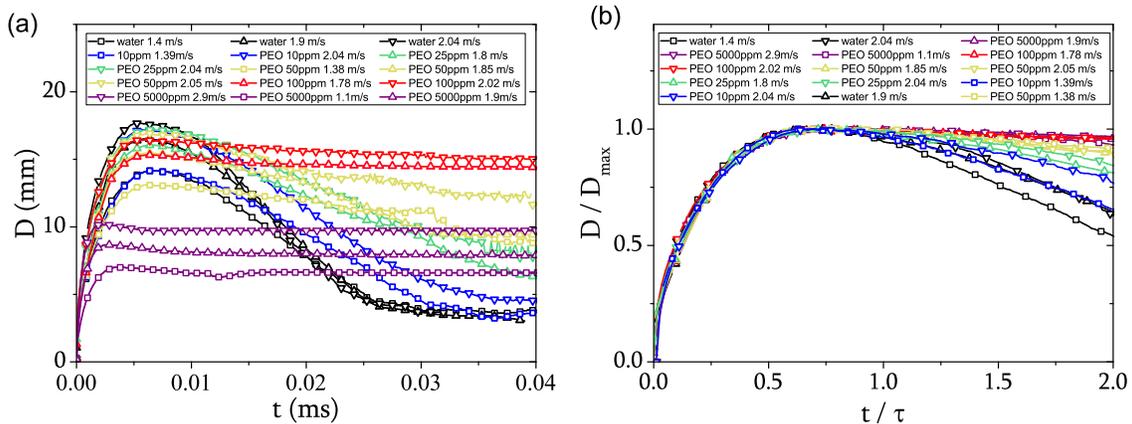}
    \caption{(a) Spreading dynamics of droplets of aqueous polymer solutions (PEO) at different concentrations and impact velocities. Water and PEO aqueous solutions at 10ppm, 25ppm, 50ppm and 100ppm are drop impacts on hydrophobic parafilm surfaces with a drop size of $D_0 = 4 \; mm$ 
    ($5200<Re<8000$ and $94<We<222$). Drop size of PEO 5000ppm is $D_0=3 \; mm$ and spreads on hydrophobic glass ($150<Re<450$ and $41<We<375$). The viscosity used is taken at shear rates given close to $V_0/D_0$ (b) Rescaled spreading dynamics for all the droplets in a).}
    \label{fig5}
\end{figure}

A surprising result comes from additional experiments carried out using polymer solutions. These solutions are non-Newtonian and show a shear thinning behaviour as shown in figure \ref{fig2}. The scaling function found for the case of pure water is not influenced by the presence of polymers despite the fact that such polymer solutions can present strong shear thinning behaviour and complex polymer flow interactions.
This is illustrated in figure \ref{fig5}a for different polymer concentrations as well as different impact velocities.
The spreading dynamics shows the usual features of increased maximum diameter and increased spreading speed versus impact velocity but also a retraction phase depending on the velocity and polymer concentration \cite{bartolo_retraction_2005} . Figure \ref{fig5}b shows the rescaled data and the very good agreement with the scaling function for water. The spreading dynamics of Newtonian and non-Newtonian fluids can be described by the same universal spreading curve. 

\begin{figure}
     \centering
    \includegraphics[width=\textwidth]{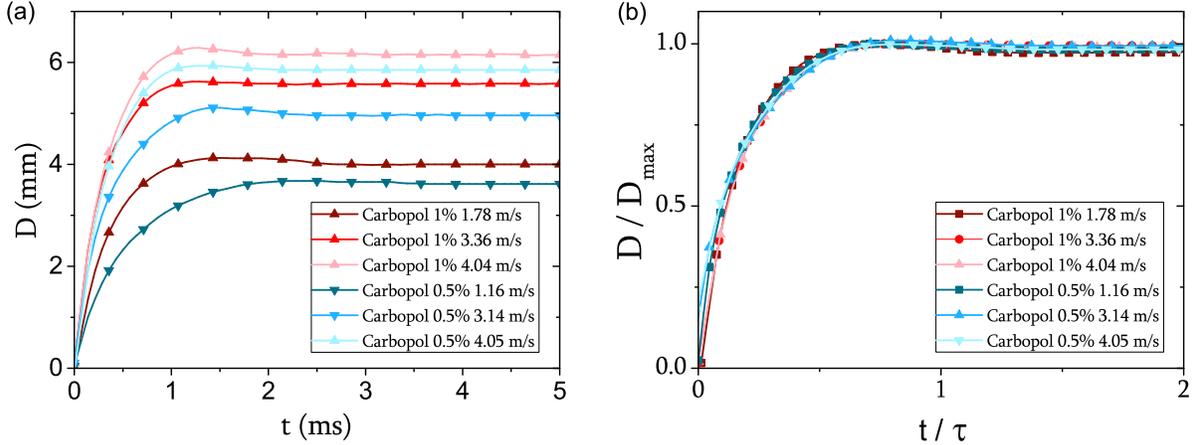}
    \caption{(a) Spreading dynamics of drop impacts of gel solutions made of water and Carbopol with two different concentrations ($0.5\%$ and $1\%$). Drop size is $D_0 = 2.3 mm$ and surface used is a hydrophilic microscope glass  ($46<We<736$ and $5<Re<31$ and $3<De<12$). The viscosity used is taken at shear rates given close to $V_0/D_0$ while the surface tension used is that of water.  (b) Rescaled spreading dynamics for all droplets in a).}
    \label{fig6}
\end{figure}

Besides the use of polymer solutions, we have carried out experiments using Carbopol solutions of different concentrations. The purpose of these experiments is to explore the role of the presence of a yield stress on the spreading dynamics. Our results are shown in Figure \ref{fig6}a. 
In all experiments, the inertial pressure defined as $\rho {V_0}^2$ which is of order $10^3$ or greater is systematically greater than the yield stress which is of order 10.
Again, a proper rescaling of the data shows the presence of a universal collapse (see figure \ref{fig6}b)
of the dynamics of the spreading up to $D_{max}$
for different impact velocities and different concentrations  and in excellent agreement with the functional form for the spreading of water droplets. 

 \makeatletter
\let\@float@original\@float
\xpatchcmd{\@float}{\csname fps@#1\endcsname}{ht!}{}{}
\makeatother
\makeatletter
\let\@float\@float@original
\makeatother

 \makeatletter
\let\@float@original\@float
\xpatchcmd{\@float}{\csname fps@#1\endcsname}{ht!}{}{}
\makeatother
\makeatletter
\let\@float\@float@original
\makeatother

\begin{figure}
     \centering
    \includegraphics[width=0.9\textwidth]{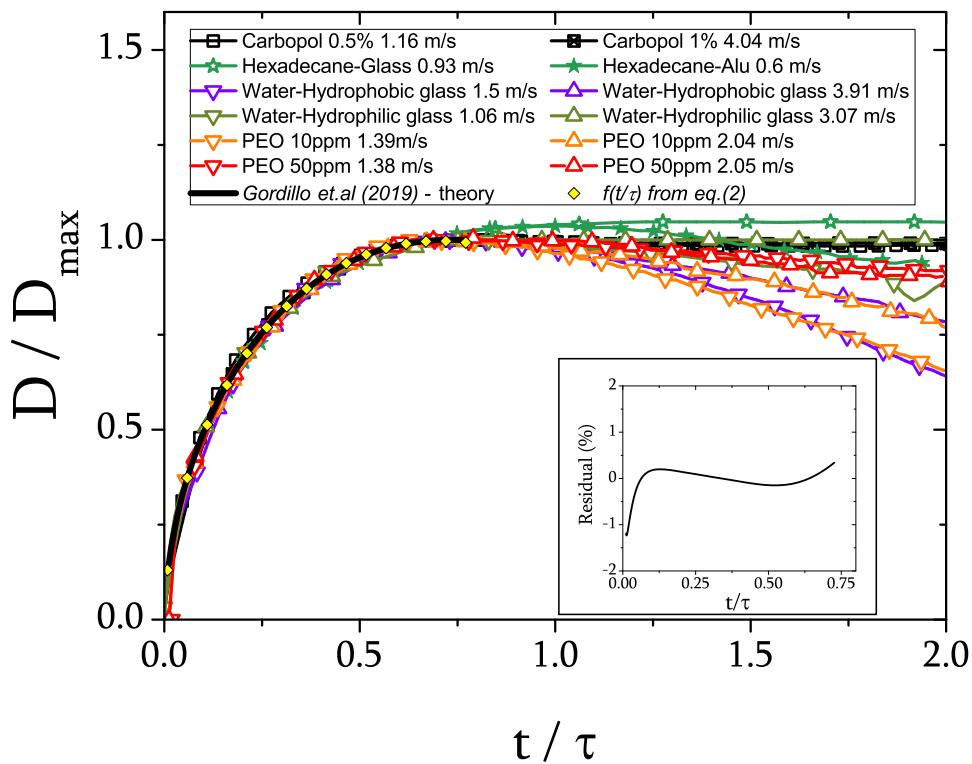}
    \caption{Rescaled spreading diameter of droplet impacts using different liquids, surfaces, drop size or impact velocities. Surface and droplet size $D_0$ used for each curves are the follow : Carbopol with hydrophilic glass and $D_0 = 2.28 mm$. Hexadecane with hydrophilic glass and $D_0 = 1.90 mm$. Water with hydrophilic and hydrophobic glass and  $D_0 = 2.04 mm$. PEO with hydrophobic parafilm and $D_0 = 4.0 mm$. The inset shows the difference between the polynomial fit and the full Gordillo solution. }
    \label{fig7}
\end{figure}

All the data obtained here follow the same scaling law and we show in figure \ref{fig7}, data from water, polymer solutions and Carbopol solutions rescaled by $D_{max}$ and $\tau=D_{max}/V_0$. An excellent collapse of all of our data from different fluids and substrates emerges showing that the dynamics of spreading, independently of rheology and substrate properties, follows a universal functional form when rescaled properly.

The rescaling of the data as proposed here shows that the spreading phase of impacting droplets follows a simple scaling function $\frac{D(t)}{D_{max}}=f(t/\tau)$. This function depends only on $D_{max}$ and $V_0$. We find that this scaling function is in excellent agreement with the solution proposed by Gordillo et.al\cite{gordillo_theory_2019}. 
Indeed, the proposed solution $D(t)$ can be found by solving a system of equations using Matlab ODE45 solver : 
\begin{equation}
    \alpha \frac{\pi}{4}\frac{db^2}{dt}=[u(s,t)-v]h(s,t)
\end{equation}
\begin{equation}
        \alpha \frac{\pi b^2}{4}\frac{dv}{dt}=[u(s,t)-v]^2 h(s,t)-(1+\beta)We^{-1}
\end{equation}
$b(t)$ is the thickness of the rim, $s(t)$ the spreading diameter, $h(s,t)$ and $u(s,t)$ are respectively the thickness and the average velocity of the thin film between the droplet and the rim, $v=\frac{ds}{dt}$ the spreading velocity of the rim and $\alpha$ and $\beta$ are two constants which are defined by the hydrophobicity of the substrate and the dynamic contact angle respectively. $\alpha=1$ when substrate is hydrophobic so the rim is lifted and thus considered with a circular shape and $\alpha=1/2$ in case of hydrophilic surface. In our case, $\alpha=1/2$ and $\beta=0$.  The results of this model are shown in figure \ref{fig7}

We have solved the system of equations proposed and rescaled the solution of the coupled equations in the same way as the experiments. The agreement between the theory of Gordillo et.al \cite{gordillo_theory_2019} , and our results shown in figure \ref{fig7} is excellent.

 \makeatletter
\let\@float@original\@float
\xpatchcmd{\@float}{\csname fps@#1\endcsname}{ht!}{}{}
\makeatother
\makeatletter
\let\@float\@float@original
\makeatother

This master curve can be used to predict the spreading dynamics of any droplet impact. While the impact velocity is a known parameter, the maximum spreading diameter $D_{max}$ needs to be found or measured. Simple expressions for $D_{max}$ such as the one proposed by Laan et.al \cite{laan_maximum_2014,lee_modeling_2016} can be used to find this value. This expression relates the maximum diameter to the Weber and Reynolds numbers of the droplet defined by $Re=\frac{\rho V_0 L}{\eta}$ and $We= \frac{\rho {V_0}^2 L}{\gamma}$ and is given by the following relation :

\begin{equation}
    \frac{D_{max}}{D_0}=\frac{\sqrt{P}}{A+\sqrt{P}}Re^{1/5}
\end{equation}

with the impact number $P=WeRe^{-2/5}$ and a fitting constant $A=1.24$.

Once the impact velocity and maximum diameter are known, our scaling function can then be used to deduce the diameter at any instant of time. For convenience and ease of use $f(t/\tau)$ can be approximated reliably by a simple a polynomial function of the form :

\begin{equation}
    f(t/\tau)=a(t/\tau)^{1/2}+b(t/\tau)+c(t/\tau)^{3/2}
\end{equation}

The constants $a$, $b$ and $c$ take the values : $a=1.401 \pm 0.002$, $b=0.903 \pm 0.008$, $c=-1.379 \pm 0.007$. 
This functional form starts out as $t^{1/2}$ at early times in accord with previous observations \cite{biance_first_2004,eddi_short_2013} with corrections at later time in powers of $t^{1/2}$.
The difference between this polynomial fit and the full solution is shown in the inset of figure \ref{fig7} where a deviation of less than a percent is observed.

The rescaling we propose as well as the agreement found with the Gordillo model has, necessarily, some limitations. Rescaling works only for drop impacts at sufficiently high velocities for which the effects of surface wettability can be neglected compared to inertia. We actually kept impact velocities sufficiently high so that droplets spread engendering a liquid film with a rim at its edge in most cases as required in the theory of Gordillo et.al. For the non Newtonian cases, the Deborah number and the impact velocities were kept large enough so that the drops can be considered as fluid by assuring that the Deborah number is much greater than 1 and that the inertial pressure is systematically larger than the yield stress. Despite these limitations, the rescaling works remarkably well up to droplet the maximum spreading of the droplets. Further, we anticipate that the proposed rescaling could be useful in other cases. Droplets impacting superheated surfaces leading to Leidenfrost drop impacts\cite{tran_drop_2012,staat_phase_2015} or droplets impacting super-hydrophobic substrates were not studied, but it is possible that the proposed rescaling may also be useful in such exotic cases.

\section{Conclusion}

We have provided a simple method to predict the spreading diameter $D(t)$ of a droplet impacting a flat surface. Spreading dynamics can be described by a universal curve for droplet impacts of Newtonian and non Newtonian fluids. This master curve is in addition in excellent agreement with the results of a theoretical model proposed recently Gordillo et.al\cite{gordillo_theory_2019} . A closed functional shape is suggested in the form of a polynomial function to approximate the spreading dynamics for any impact velocity, droplet size, different substrates and fluid types.

\bibliography{bibliography}

\newpage

\end{document}